\documentclass[
  aip,
  floatfix,
  jcp,
  twoside,
  twocolumn,
  10pt,
  showpacs,
  citeautoscript,
  superscriptaddress,
]{revtex4-2}

\usepackage{amsmath}
\usepackage{amssymb}
\usepackage{booktabs}
\usepackage{braket}
\usepackage{float}
\usepackage{glossaries}
\usepackage{graphicx}
\usepackage{tabularx}
\usepackage{units}
\usepackage[version=4]{mhchem}
\usepackage[dvipsnames]{xcolor}
\usepackage[colorlinks=true,
            linkcolor=NavyBlue,
            citecolor=NavyBlue,
            filecolor=NavyBlue,
            urlcolor=NavyBlue
            ]{hyperref}

\graphicspath{{./figures/}}

\renewcommand{\vec}[1]{\ensuremath\boldsymbol{#1}}
\newcommand{\cmat}[1]{\ensuremath\underline{\boldsymbol{#1}}}
\newcommand{\gpaw}{\textsc{gpaw}}
\newcommand{\nwchem}{\textsc{NWChem}}

\setacronymstyle{long-short}
\newacronym{dc}{DC}{dipolar coupling}
\newacronym{dft}{DFT}{density functional theory}
\newacronym{dos}{DOS}{density of states}
\newacronym{homo}{HOMO}{highest occupied molecular orbital}
\newacronym{ks}{KS}{Kohn-Sham}
\newacronym{lcao}{LCAO}{linear combination of atomic orbitals}
\newacronym{lrtddft}{LR-TDDFT}{linear-response TDDFT}
\newacronym{lumo}{LUMO}{lowest unoccupied molecular orbital}
\newacronym{np}{NP}{nanoparticle}
\newacronym{rttddft}{RT-TDDFT}{real-time TDDFT}
\newacronym{sc}{SC}{strong coupling}
\newacronym{si}{SI}{Supplementary Information}
\newacronym{tddft}{TDDFT}{time-dependent DFT}
\newacronym{xc}{XC}{exchange correlation}

\newcommand{\phys}{
    Department of Physics,
    Chalmers University of Technology,
    SE-412~96 Gothenburg, Sweden
}
\newcommand{\aalto}{
    Department of Applied Physics,
    Aalto University,
    00076 Aalto, Finland
}
\newcommand{\jyvaskyla}{
    Department of Chemistry,
    University of Jyväskylä,
    40014 Jyväskylä, Finland
}
\newcommand{\warsaw}{
    Faculty of Physics,
    University of Warsaw,
    02-093 Warsaw, Poland
}

\begin{document}

\title{
    Dipolar coupling of nanoparticle-molecule assemblies:
    An efficient approach for studying strong coupling
}

\author{Jakub\ Fojt}
\affiliation{\phys}

\author{Tuomas\ P.\ Rossi}
\affiliation{\aalto}

\author{Tomasz J.\ Antosiewicz}
\affiliation{\warsaw}

\author{Mikael\ Kuisma}
\affiliation{\jyvaskyla}

\author{Paul\ Erhart}
\email[Corresponding author: ]{erhart@chalmers.se}
\affiliation{\phys}

\begin{abstract}
Strong light-matter interactions facilitate not only emerging applications in quantum and non-linear optics but also modifications of materials properties.
In particular the latter possibility has spurred the development of advanced theoretical techniques that can accurately capture both quantum optical and quantum chemical degrees of freedom.
These methods are, however, computationally very demanding, which limits their application range.
Here, we demonstrate that the optical spectra of nanoparticle-molecule assemblies, including strong coupling effects, can be predicted with good accuracy using a subsystem approach, in which the response functions of the different units are coupled only at the dipolar level.
We demonstrate this approach by comparison with previous time-dependent density functional theory calculations for fully coupled systems of Al nanoparticles and benzene molecules.
While the present study only considers few-particle systems, the approach can be readily extended to much larger systems and to include explicit optical-cavity modes.
\end{abstract}

\maketitle

\section{Introduction}

The coupling of light and matter in the \gls{sc} regime leads to the emergence of excited states of mixed nature \cite{Ebb16}, which are characterized by a coherent energy exchange between the subsystems at a rate that is much faster than the respective damping rates.
Emitter and the cavity thus form a light-matter hybrid (polariton) with modified and tunable properties \cite{TorBar15, BarWerCua18, FriMirDeL19}, including nonlinear and quantum optical phenomena \cite{YosSchHen04, BirBocMil05, KasRicKun06}, photochemical rates \cite{HutSchGen12, HerSpa16, GalGarFei16, MarRibCam18, MunWerBar18, FliNar20, FreGraPer20}, thermally-activated ground-state chemical reactions under vibrational \gls{sc} \cite{ThoGeoSha16, ThoLetNag19, HirTakRie20} as well as exciton transport \cite{FeiGar15, ZhoCheWan16}.

Theoretical analysis of these phenomena is non-trivial as polaritons reside at the intersection between quantum optics, quantum chemistry, and solid state physics.
While quantum optical approaches such as Jaynes--Cummings or Dicke models have been used extensively \cite{JayCum63, Gar11, TorBar15, BarWerCua18, FriMirDeL19}, they are ill-suited for describing the material aspects.
This has spurred the development of advanced theoretical techniques in recent years based on various quantum optical and quantum chemistry methods \cite{HerSpa16, GalGarFei16, MarRibCam18, GalGarFei15, RugFliPel14, FliRugApp17, LinFeiTop17, NeuEstCas18, RosSheErh19, FliNar20, FliWelRug19, FreGraPer20}.
While those methods that provide an accurate account of quantum chemical effects are computationally very demanding, methods with a more approximate treatment of the materials degrees of freedom are limited with regard to chemical specificity.
At the same time, ensemble effects and the collective interaction of, e.g., molecules and/or \glspl{np} are of immediate experimental interest \cite{OrgGeoHut15, ThoGeoSha16}, calling for methods that allow one to bridge between system specific predictions and computational efficiency.

We have recently demonstrated the usefulness of \gls{dft} \cite{HohKoh64, KohSha65} and \gls{tddft} \cite{RunGro84} calculations for studying polariton physics in \gls{np}-molecule systems \cite{RosSheErh19}.
Here, building on this study, we demonstrate that the observed effects can be reproduced nearly quantitatively using a subsystem approach where the different units, specifically \glspl{np} and molecules, only interact via \gls{dc}.
This approach allows one to very quickly evaluate the coupled response for a wide range of geometries at a computational cost that is orders of magnitudes smaller than a full \gls{tddft} approach.
In addition, it enables one to combine response function calculations from a variety of sources, including, e.g., different first-principles techniques, exchange-correlation functionals, and classical electrodynamic calculations.

In the following section, we provide an overview of the methodology used in this study as well as computational details.
We then consider different scenarios, including (\emph{i}) coupling as a function of \gls{np}-molecule distance (\autoref{sect:results:distance-dependence}), (\emph{ii}) coupling as a function of the number of molecules and the size of the \gls{np} (\autoref{sect:results:number-of-molecules}), and (\emph{iii}) mixing dynamic polarizabilities from different sources and/or types of calculations (\autoref{sect:results:mixing-sources}).
Finally, we summarize the conclusions and provide an outlook with respect to anticipated improvements and extensions.

\section{Methodology}
\label{sec:methods}

In the dipolar coupling limit, the response of a system to an electrical field can be described by its dynamic polarizability tensor $\alpha_{\mu\nu}(\omega)$.
In the following sections, we review the dynamic polarizability and the coupled response of a set of polarizable units within this framework.

\subsection{Dynamic polarizability}
\label{sec:methods:polarizability}

The dynamic polarizability $\vec \alpha(t)$ is a linear response function that defines the induced dipole moment $\vec d(t)$ in response to the external field $\vec{E}_\text{ext}(t)$,
\begin{align}
    d_\mu(t)
    &=
    \int_{-\infty}^{\infty}\sum_{\nu} \alpha_{\mu\nu}(t-\tau) E^\text{ext}_\nu(\tau) {\rm d}\tau .
    \label{eq:induced-dipole-moment}
\end{align}
Here and in the following, $\mu$ and $\nu$ refer to the Cartesian coordinates of the response and external electric field respectively.
We emphasize that $\vec \alpha(t)$ is a causal response function that is zero in the negative time domain.

In frequency space the dipole response is obtained via Fourier transformation and convolution theorem as
\begin{align}
    \label{eq:alpha}
    d_\mu(\omega) &= \sum_\nu \alpha_{\mu\nu}(\omega) E^\text{ext}_\nu(\omega).
\end{align}
Hence, the $\alpha_{\mu\nu}$ component of the polarizability tensor describes the strength of the response along $\mu$ given a field along $\nu$.

In practice, a typical calculation of the polarizability tensor within \gls{rttddft} consists of recording the induced time-dependent dipole moment following a weak electric field impulse \cite{YabBer96}.
Then, the response in frequency space is obtained by Fourier transformation
\begin{align}
    d_\mu(\omega) &= \int_{0}^{\infty} d_\mu(t) e^{i \omega t} {\rm d}t.
    \label{eq:fourier}
\end{align}
By choosing the external electric field $E^\text{ext}_\nu(t) = K_0  \delta(t)$ to be aligned along $\nu$, the $\mu\nu$ component of the polarizability tensor is obtained as
\begin{align}
    \alpha_{\mu\nu}(\omega) = \frac{d_\mu(\omega)}{K_0}.
\end{align}
The full polarizability tensor can thus be calculated by carrying out at most three calculations with external fields along each Cartesian direction; this number can be reduced further in the presence of symmetry.
Here, the strength of the impulse, $K_0$, is set to be sufficiently weak so that the full response given by \gls{rttddft} is dominated by the linear response regime.

The time-propagation approach allows calculating metallic nanoparticles with hundreds of atoms \cite{KuiSakRos15} in particular when combined with \gls{xc} functionals that can accurately account for the $d$-band position in noble metal \glspl{np}.

For completeness, we note that for molecules it is usually more convenient to evaluate the excitation spectrum directly in the frequency domain using the Casida approach for \gls{lrtddft} \cite{Cas95, Cas09}.
In this case the response can be formally obtained by solving a matrix eigenvalue equation, which yields the excitation energies $\omega_I$ and the corresponding transition dipole moments $\braket{\Psi_I| r_\mu | \Psi_0}$, and the dynamic polarizability can then be obtained as
\begin{align}
    \alpha_{\mu\nu}(\omega) = \sum_I \frac{2 \omega_I \braket{\Psi_0| r_\mu |\Psi_I}\braket{\Psi_I| r_\nu |\Psi_0}}{\omega_I^2 - \omega^2}.
    \label{eq:alpha:casida}
\end{align}

To deal with the divergence along the real frequency axis and to obtain (artificial) broadening of the spectrum, it is common to set $\omega \to \omega + i \eta$ both in Eqs.~\eqref{eq:fourier} and \eqref{eq:alpha:casida}, resulting in a Lorentzian line shape with a peak width that is determined by the $\eta$ parameter.

Below, we also consider coupling to a homogeneous sphere described by a local dielectric function $\varepsilon(\omega)$ according to Mie theory\cite{Mie08}.
To this end, we note that in the quasi-static limit the dynamic polarizability tensor of a sphere with volume $V$ is diagonal and takes the form\cite{Sih07}
\begin{equation}
    \alpha_{\mu\nu}(\omega) = 3\varepsilon_0 V \frac{\varepsilon(\omega) - 1}{\varepsilon(\omega) + 2}\delta_{\mu\nu}.
\end{equation}

\subsection{Dipolar coupling}
\label{sec:methods:coupling}

Given the dynamic polarizability of two or more units we can evaluate the response of the coupled system.
This is a widely known approach for molecular assemblies, see for example Refs.~\citenum{DeV64, FidKnoWie91, AugDarRu19}, but we present it here in whole for completeness.

\begin{widetext}
Let $\vec\alpha^{(i)}_0(\omega)$ be the irreducible polarizability tensor of the individual units.
For a system composed of $N$ units, the induced dipole moments, given the total electric field  $\vec E_\text{tot}^{(i)}(\omega)$ at each unit, are obtained as
\begin{align}
    \underbrace{
    \begin{bmatrix}
        \vec d^{(1)}(\omega) \\
        \vec d^{(2)}(\omega) \\
        \vdots \\
        \vec d^{(N)}(\omega) \\
    \end{bmatrix}
    }_{\textstyle\cmat d(\omega)}
    =
    \underbrace{
    \begin{bmatrix}
        \vec\alpha_0^{(1)}(\omega) & & & \\
        & \vec\alpha_0^{(2)}(\omega) & & \\
        &   & \ddots  & \\
        & & & \vec\alpha_0^{(N)}(\omega)
    \end{bmatrix}
    }_{\textstyle\cmat\alpha_0(\omega)}
    \underbrace{
    \begin{bmatrix}
        \vec E_\text{tot}^{(1)}(\omega) \\
        \vec E_\text{tot}^{(2)}(\omega) \\
        \vdots \\
        \vec E_\text{tot}^{(N)}(\omega) \\
    \end{bmatrix}
    }_{\textstyle\cmat E_\text{tot}(\omega)}
    \label{eq:mu-ind}
    .
\end{align}

The Coulomb interaction between units $i$ and $j$, located at $\vec R_i$ and $\vec R_j$ is given in atomic units by $1 / |\vec R_{ij}|$, where $\vec R_{ij} = \vec R_i-\vec R_j$.
The dipole-dipole interaction between point dipoles at $\vec R_i$ and $\vec R_j$ is given by a tensor
\begin{equation}
T_{i\mu j\nu} = \nabla_{\mu} \nabla_{\nu} \frac{1}{|\vec R_{ij}|} =
    \frac{ \delta_{\mu\nu}}{|\vec R_{ij}|^{3}}
    - 3 \frac{R_{ij,\mu} R_{ij,\nu}}{|\vec R_{ij}|^5}.
\end{equation}
Therefore, the total electric field at each polarizable unit is obtained as
\begin{align}
    \underbrace{
    \begin{bmatrix}
        \vec E_\text{tot}^{(1)}(\omega) \\
        \vec E_\text{tot}^{(2)}(\omega) \\
        \vdots  \\
        \vec E_\text{tot}^{(N)}(\omega) \\
    \end{bmatrix}
    }_{\textstyle\cmat E_\text{tot}(\omega)}
    =
    \underbrace{
    \begin{bmatrix}
        \vec E_\text{ext}^{(1)}(\omega) \\
        \vec E_\text{ext}^{(2)}(\omega) \\
        \vdots \\
        \vec E_\text{ext}^{(N)}(\omega) \\
    \end{bmatrix}
    }_{\textstyle \cmat E_\text{ext}(\omega)}
    -
    \underbrace{
    \begin{bmatrix}
        0 & \vec T_{12} & \ldots &  \vec T_{1N} \\
        \vec T_{21} &    0   & & \vec T_{2N} \\
        \vdots &   & \ddots  & \vdots \\
        \vec T_{N1} & \vec T_{N2} & \ldots & 0
    \end{bmatrix}
    }_{\textstyle \cmat T}
    \underbrace{
    \begin{bmatrix}
        \vec d^{(1)}(\omega) \\
        \vec d^{(2)}(\omega) \\
        \vdots \\
        \vec d^{(N)}(\omega) \\
    \end{bmatrix}
    }_{\textstyle\cmat d(\omega)}
    .
    \label{eq:E-tot}
\end{align}
\end{widetext}

By substituting Eq.~\eqref{eq:E-tot} into Eq.~\eqref{eq:mu-ind} and solving for the induced dipole moment we obtain
\begin{align}
    \cmat d(\omega)
    &=
    \left[\cmat I + \cmat\alpha_0(\omega) \cmat T \right]^{-1} \cmat \alpha_0(\omega) \cmat E_\text{ext}(\omega),
\end{align}
where the reducible unit-wise polarizability tensor is identified
\begin{subequations}
\begin{align}
    \cmat \alpha(\omega)
    &= \left[\cmat I + \cmat\alpha_0(\omega) \cmat T \right]^{-1} \cmat \alpha_0(\omega) \\
    &= \left[\cmat\alpha_0(\omega)^{-1} + \cmat T\right]^{-1}.
\end{align}
\end{subequations}
Each unit contributes to the total dipole moment of the coupled system.
Assuming a uniform external electric field throughout the coupled system ($\vec E_\text{ext}^{(1)} = \vec E_\text{ext}^{(2)} = \ldots = \vec E_\text{ext}^{(N)}$), the total polarizability tensor for coupled system comprising $N$ units is obtained by carrying out a double summation over all units of the system
\begin{align}
    \alpha_{\mu\nu}(\omega) = \sum_{i}^{N} \sum_{j}^{N} [\cmat \alpha]_{i\mu j \nu}(\omega).
    \label{eq:alpha:coupled}
\end{align}

To present the results, we use the dipole strength function given by the imaginary part of the dynamic polarizability
\begin{align}
    S_\mu(\omega) = \frac{2 \omega}{\pi} \Im[\alpha_{\mu\mu}(\omega)].
\end{align}
The dipole strength function equals the electronic photoabsorption spectrum safe for a constant multiplier and satisfies the Thomas--Reiche--Kuhn sum rule analogously to the oscillator strength.

\subsection{Computational details}
\label{sect:computational-details}

The coupled systems considered in this study comprise one Al \gls{np} with 201, 586 or 1289 atoms and one or more benzene molecules, which have been investigated in whole using \gls{rttddft} calculations in Ref.~\citenum{RosSheErh19}.
The formalism outlined in the previous section requires material specific input in the form of the dynamic polarizability of the individual subsystems, i.e. isolated Al \glspl{np} of different sizes and an isolated benzene molecule, respectively.

The data for these systems have been calculated in Refs.~\onlinecite{RosSheErh19, RosSheErh19Data} using the PBE \gls{xc} functional \cite{PerBurErn96a} in the adiabatic limit and \gls{rttddft} via the $\delta$-kick technique \cite{YabBer96} (\autoref{sec:methods:polarizability}) as implemented using \gls{lcao} basis sets \cite{KuiSakRos15} in the \gpaw{} code. \cite{EnkRosMor10}
The projector augmented-wave \cite{Blo94} method was employed with double-$\zeta$ polarized (dzp) basis sets as provided in \gpaw{}.
The wave functions were propagated up to \unit[30]{fs} using a time step of \unit[15]{as}.
Further details of these calculations are given in Refs.~\onlinecite{RosSheErh19, RosSheErh19Data}.

For benzene we also evaluated the first 16 roots of the excitation spectrum using \gls{lrtddft} within the Casida approach \cite{Cas95, Cas09} as implemented in the \nwchem{} suite \cite{ValBylGov10}.
Calculations were carried out using the B3LYP functional \cite{LeeYanPar88, Bec93} and the 6-311G$^*$ basis set \cite{McLCha80, KriBinSee80}.
The excitation spectra were subsequently transformed into dynamic polarizabilities via Eq.~\eqref{eq:alpha:casida}.

All spectra obtained in this work or taken from Ref.~\citenum{RosSheErh19Data} are broadened using $\eta=\unit[0.1]{eV}$.
For the purpose of extracting coupling strength parameters, a coupled oscillator model \cite{WuGraPel10} was fitted to the obtained spectra.
The details of the fitting scheme are outlined in Supplementary Note~\ref*{snote:fit}.

\section{Results and discussion}

\subsection{Dynamic polarizability of individual \texorpdfstring{\glspl{np}}{NP} and molecules}
\label{sect:results:dynamic-polarizabilities}

\begin{figure}
    \centering
    \includegraphics{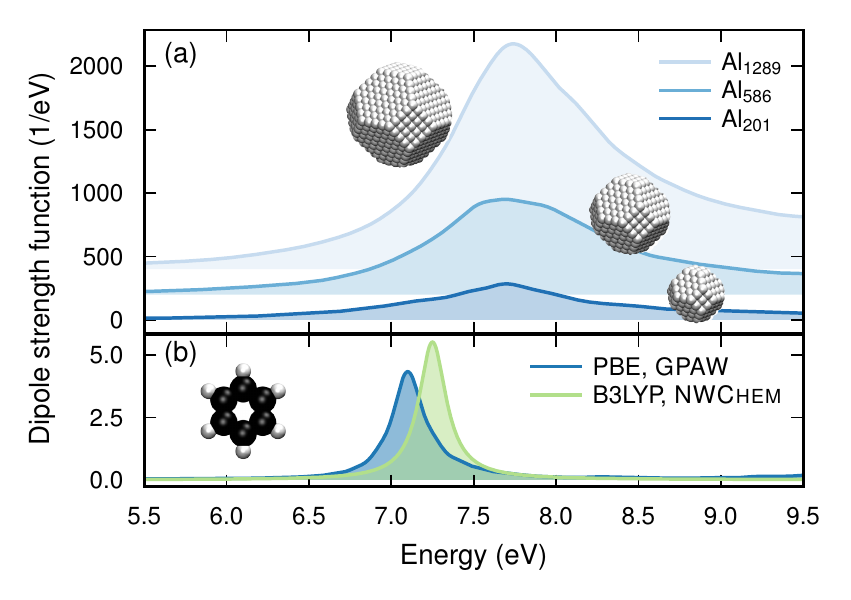}
    \caption{
        Optical spectra for isolated (a) Al \glspl{np} and (b) benzene molecules.
        All \gls{np} data in (a) were obtained from \gpaw{} (\gls{rttddft}) using the PBE \gls{xc} functional.
        The benzene data in (b) were obtained from \gpaw{} (\gls{rttddft}) and \nwchem{} (Casida) calculations using different \gls{xc} functionals as indicated in the legend.
    }
    \label{fig:isolated-units}
\end{figure}

In the following we consider Al \glspl{np} with 201, 586 and 1289 atoms, which exhibit a plasmon resonance at about \unit[7.7]{eV} (\autoref{fig:isolated-units}a).
Below we analyze their coupling to two or more benzene molecules.
The response of the latter is obtained either at the PBE or the B3LYP level yielding the first bright excitation at \unit[7.1]{eV} and \unit[7.3]{eV}, respectively (\autoref{fig:isolated-units}b).

\subsection{Coupling as a function of distance}
\label{sect:results:distance-dependence}

Using the dynamic polarizabilities of the isolated units in Eq.~\eqref{eq:alpha:coupled}, we first inspect the optical spectrum of a system comprising an \ce{Al201} \gls{np} and two benzene molecules as a function of the \gls{np}-molecule distance.
Reference data from \gls{rttddft} calculations obtained using the PBE \gls{xc} functional for the full system are available from Refs.~\onlinecite{RosSheErh19, RosSheErh19Data}.
These full \gls{tddft} calculations take into account not only coupling to all orders but also charge transfer and renormalization of the underlying states and excitations due to \gls{np}-molecule interactions.

\begin{figure*}
    \centering
    \includegraphics{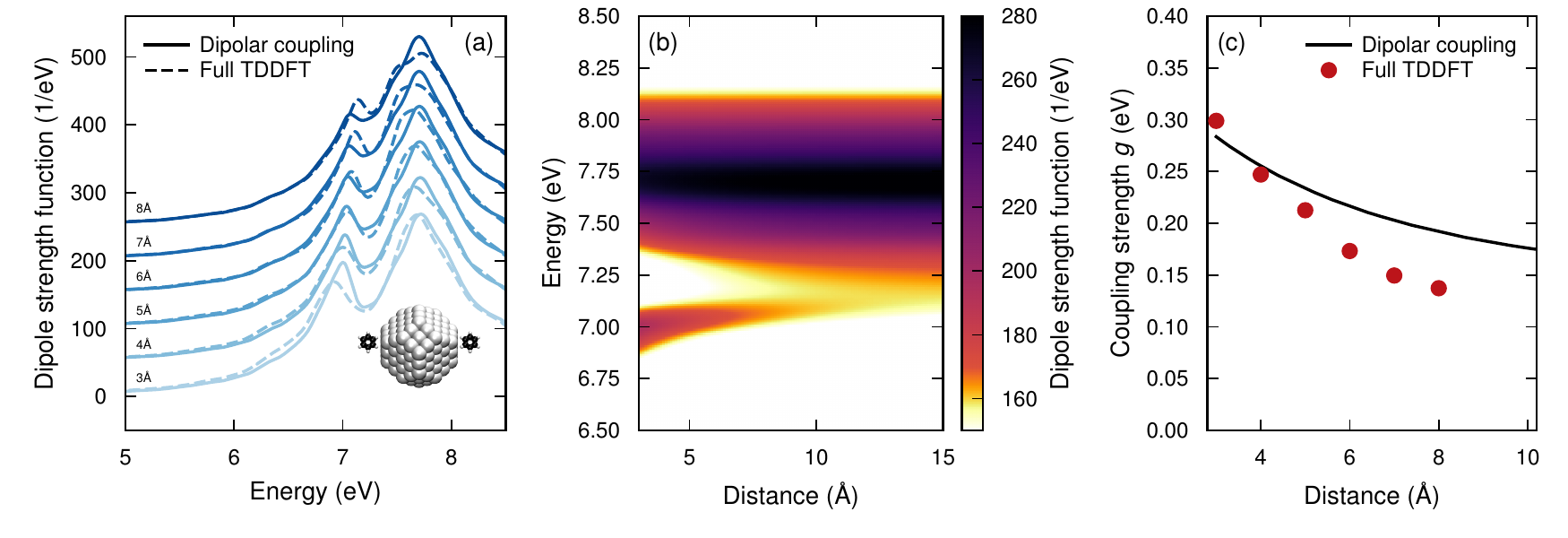}
    \caption{
        Distance dependence of the optical response from dipolar coupling and full \gls{tddft} calculations.
        (a) Optical spectra for a system comprising a \ce{Al201} \gls{np} and two benzene molecules as the \gls{np}-molecule distance is varied from \gls{rttddft} calculations for the full system (Full TDDFT) and from Eq.~\eqref{eq:alpha:coupled} (Dipolar coupling).
        (b) Spectrum of \ce{Al201}+2 benzene system as a practically continuous function of the \gls{np}-molecule distance obtained using the dipolar coupling approximation.
        (c) Coupling strength extracted from Bayesian fits of the spectra in (a) and (b) to a coupled oscillator model.
        The extracted values are shown as round markers for the full system (Full TDDFT) and as a solid line for the dipolar coupling calculations.
    }
    \label{fig:distance-dependence}
\end{figure*}

The spectra obtained in the \gls{dc} approximation and from full \gls{tddft} calculations are in good agreement over the entire range of distances considered here (\autoref{fig:distance-dependence}a).
Since the \gls{dc} calculations only include dipolar interactions, the good agreement suggests that for the system under study, higher-order terms, charge transfer, and orbital hybridization play a rather small role in the part of the configuration space considered here.

For the shortest distance of \unit[3]{\AA} a slight underestimation becomes apparent of both position and width of the lower polariton, which we attribute to the absence of orbital overlap and to a lesser extent the neglect of higher-order multipole interactions.
In the full \gls{tddft} calculation, the former contribution, which is the more dominant one owing to the fact that higher order multipoles should be even more blue detuned from the benzene transition than the dipolar one,  results in a broadening of the benzene transition, which couples with rate $g$ to the Al \gls{np} plasmon.
An increase of the decay rate ($\gamma$) of one of the strongly coupled elements causes the Rabi splitting $\Omega$ to increase as $\Omega=\sqrt{4g^2 -(\gamma_\mathrm{pl}-\gamma_\mathrm{ex})^2}$ (for zero detuning) \cite{BarWerCua18}, causing the \gls{tddft}-computed polaritons to be broader and farther apart than in the calculations using the \gls{dc} approximation.
Also, since the Al-\gls{np} plasmon is blue-detuned from the benzene transition, the two subsystems do not contribute equally to both polaritons.
The lower polariton has a more benzene-like character and is more influenced by the width of the benzene transition.
Conversely, the upper polariton is more Al-plasmon-like and is not influenced significantly by the decay rate of the molecular transition.

In spite of the limitations discussed in the previous paragraph, the clear strength and benefit of the \gls{dc} calculations lies in their very rapid computation.
Indeed, once the response functions of the individual units have been computed, it is straightforward (and orders of magnitude faster than with full \gls{tddft} modeling) to map out the configuration space.
The \gls{dc} calculations of hybrid systems can even have certain advantages compared to full \gls{tddft} calculations using incomplete basis sets, as numerical factors that can affect the accuracy of the latter are effectively avoided.

Following the evolution of the spectrum as a practically continuous function of the \gls{np}-molecule distance reveals the emergence of a clear lower polariton state starting at distances of about \unit[10]{\AA} (\autoref{fig:distance-dependence}b).
The coupling strengths $g$ extracted from the \gls{dc} calculations agree well with the full \gls{tddft} calculations, both with respect to magnitude and distance dependence (\autoref{fig:distance-dependence}c).
Interestingly one observes the difference in $g$ between the two types of calculations to increase with distance, which might appear unintuitive at first.
This increasing difference stems from the different sources of error in the \gls{tddft} and \gls{dc} calculations.
In the former, the use of localized basis sets that shift in space between calculations induces numerical errors, which exaggerate the blue-shift of the lower polariton at large distances (Supplementary Fig.~\ref*{sfig:params}).
In the latter, the missing effect of state hybridization leads to overly pronounced features, i.e. the polariton peaks are higher and the valley between them deeper.
In the coupled oscillator model, an increased value $g$ both makes the lower polariton peak more pronounced and the separation between lower and upper polariton larger.

\subsection{Coupling as a function of the number of molecules and \texorpdfstring{\gls{np}}{NP} size}
\label{sect:results:number-of-molecules}

\begin{figure}
    \centering
    \includegraphics{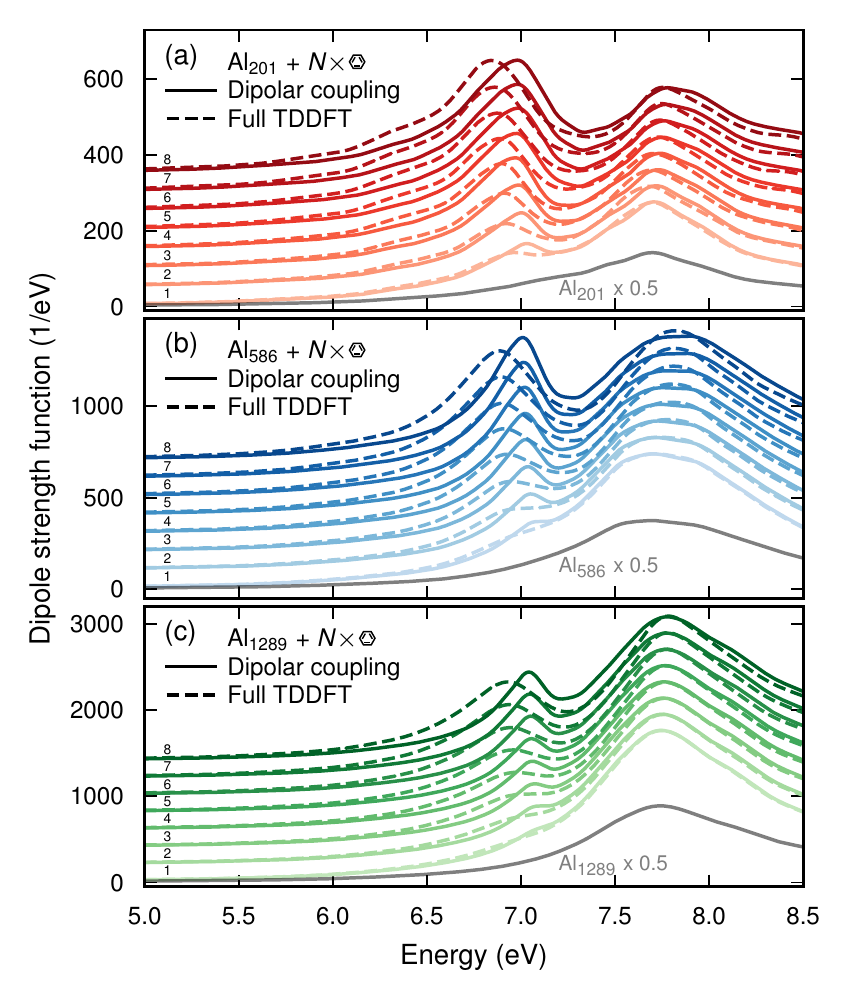}
    \caption{
        Dependence of the optical response on the number of molecules from dipolar coupling and full \gls{tddft} calculations.
        Optical spectra of a system comprising (a) a \ce{Al201} \gls{np}, (b) a \ce{Al586} or (c) a \ce{Al1289} and a variable number of benzene molecules from \gls{rttddft} calculations for the full system (Full TDDFT) and from Eq.~\eqref{eq:alpha:coupled} (Dipolar coupling).
        The solid gray lines show the spectrum of the isolated \glspl{np} for comparison.
    }
    \label{fig:number-of-molecules-dependence}
\end{figure}

Having confirmed the ability of the \gls{dc} approximation to reproduce the distance dependence for an Al \gls{np} with two benzene molecules and the concomitant emergence of strong coupling, it is now instructive to analyze the spectrum as a function of the number $N$ of benzene molecules, which provides another means for tuning the coupling.
As their number increases the effective mass of benzene excitations in the \gls{np}-molecule hybrid system increases, enhancing the coupling rate proportionally to the square root of $N$.
The resulting large coupling strengths induce much larger changes of the polariton spectra than when the \gls{np}-molecule distance is varied.
In particular, one notes that both the lower and upper polaritons exhibit significant changes with the Rabi splitting growing monotonically with $N$ \cite{RosSheErh19}.

We consider Al \glspl{np} with three sizes containing 201, 586 and 1289 atoms coupled to up to eight benzene molecules.
Again the \gls{dc} approximation works well overall, although the agreement becomes worse as \gls{np} size and/or the number of benzene molecules increases (\autoref{fig:number-of-molecules-dependence}).

For the \ce{Al201} case with several benzene molecules, one observes a gradual shift of spectral weight from the upper to the lower polariton.
For small $N$ it is sensible to attribute the lower (upper) polariton primarily benzene (Al-\gls{np}) character.
For larger $N$, however, the distinction between plasmonic (Al-\gls{np}) and excitonic character (benzene molecule) becomes invalid as the system forms a fully-mixed state (\autoref{fig:number-of-molecules-dependence}a).
This marked change between the balance of the upper and lower polaritons with $N$ is the result of a gradual shift from a red-detuned molecular transition of one benzene with respect to the \gls{np}-plasmon to a blue-detuned one for $N=8$ \cite{RosSheErh19}.
This change is mostly ascribed to the red-shift of the \gls{np}-plasmon with increasing $N$, which may not be captured quantitatively in the \gls{dc}-approximation.
The \emph{position} of the lower polariton in this \ce{Al201} sequence is slightly blue-shifted relative to the full \gls{tddft} data.
On the other hand, the \gls{dc} calculations yield rather sensible results for the \emph{width} of both polaritons.

By comparison, for the larger \ce{Al586} and \ce{Al1289} \glspl{np} one observes a systematic underestimation of \emph{both} position and width of the lower polariton (\autoref{fig:number-of-molecules-dependence}b,c).
As noted above (\autoref{sect:results:distance-dependence}), the differences between \gls{dc} and \gls{tddft} calculations at a distance of \unit[3]{\AA} are likely related to the absence of orbital hybridization in the \gls{dc} approximation.
The orbital hybridization in \gls{tddft} may also be the cause of other, potentially small, effects that escape the \gls{dc} approximation.
In addition to the above mentioned red-shift of the \gls{np}-plasmon, the presence of the molecule may itself modify the mode volume of the cavity (beyond the explicitly studied plasmon-transition coupling) and increase the coupling strength \cite{YanAntShe16a}.

\subsection{Response functions from multiple sources}
\label{sect:results:mixing-sources}

\begin{figure}
    \centering
    \includegraphics{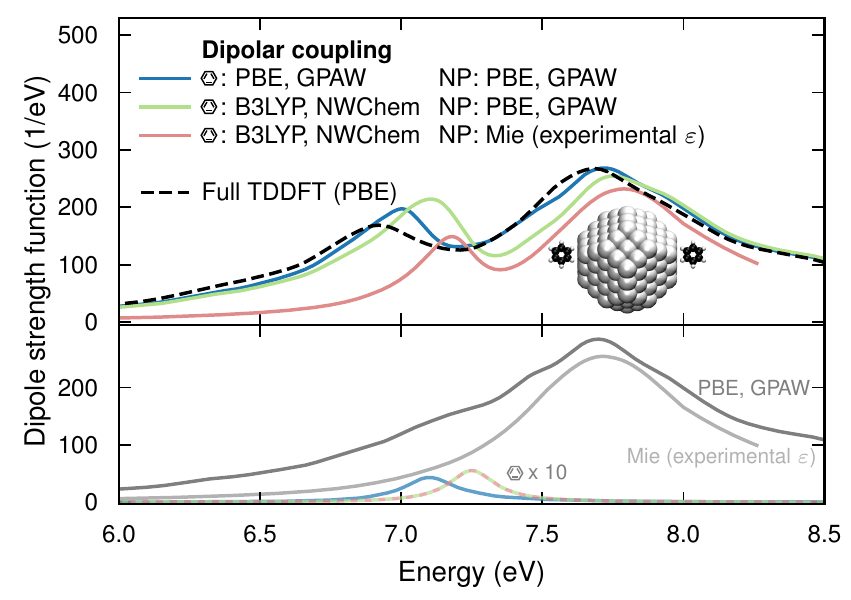}
    \caption{
        Mixing response calculations from different sources.
        Optical response of a system comprising \ce{Al201} and two benzene molecules at a distance of \unit[3]{\AA}, where the dynamic polarizability of the former is alternatively either by using the PBE \gls{xc} functional and \gls{rttddft} within the \gpaw{} code or by using the Mie approximation in the quasi-static limit with an experimental dielectric function.
        The dynamic polarizability of the latter is alternatively obtained by using the PBE \gls{xc} functional and \gls{rttddft} within the \gpaw{} code or by using the B3LYP \gls{xc} functional and Casida \gls{lrtddft} within the \nwchem{} code.
        The response of the isolated systems is shown for reference at the bottom of the figure using shaded lines.
    }
    \label{fig:pbe-vs-b3lyp}
\end{figure}

While the full \gls{tddft} calculations can capture a variety of contributions that are missed by \gls{dc} calculations, they are effectively limited by the need to describe the system using one common \gls{xc} functional.
Yet, functionals that perform well for metals are often poorly suited for molecular systems and vice versa.
A distinct advantage of the \gls{dc} approximation is its ability to combine response functions from multiple different sources, including but not limited to different types of electronic structure calculations as well as classical electrodynamics simulations.
Here, we specifically exploit this feature to analyze the effect of the \gls{xc} functional with regard to the description of the excitation spectrum of benzene.
Simultaneously, we highlight the possibility for inter-code coupling when using the \gls{dc} approximation.

We consider a system comprised of a \ce{Al201} \gls{np} and two benzene molecules at a distance of \unit[3]{\AA} using dynamic polarizabilities calculated using different \gls{xc} functionals (PBE and B3LYP) and different electronic structure codes (\gpaw{} and \nwchem{}) (\autoref{fig:pbe-vs-b3lyp}).
In comparison to PBE, the B3LYP \gls{xc} functional gives the first bright excitation higher in energy, closer to both the experimental value and the plasmon resonance of the Al \gls{np}.
The larger spectral overlap leads to less detuning, which primarily manifests itself in a more pronounced transfer of oscillator strength from the upper to the lower polariton.
As proof-of-concept, we also demonstrate the \gls{dc} approximation using the polarizability of a Mie sphere with the dielectric function of bulk Al by McPeak \textit{et al}. \cite{McPJayKre15}
By matching the diameter of the Mie sphere to the distance between opposing \{100\} facets of the \ce{Al201} \gls{np}, \unit[16.2]{Å}, the spectra are similar.
We note that the deviation between the Mie spectra based on the dielectric functions measured experimentally by McPeak \textit{et al}. and Rakic \cite{Rak95} is considerably larger than the deviation between the present calculation and the spectrum based on the data from McPeak \textit{et al.}

\section{Conclusions and outlook}

In this study we have analyzed the efficacy of the \gls{dc} approximation for capturing the emergence of \gls{sc} in Al \gls{np}-benzene hybrid systems.
To this end, we compared optical spectra obtained from \gls{dc} calculations with the results from an earlier study that applied \gls{tddft} to the entire system \cite{RosSheErh19}.
We find that overall the \gls{dc} approximation is able to reproduce the full \gls{tddft} results for this system well over a large size range of considered \glspl{np}.
Deviations become more pronounced at short \gls{np}-molecule distances as orbital overlap and higher-order multipole interactions start to play a role.

After computation of the dynamic polarizabilities of the isolated components, the \gls{dc} approximation is many orders of magnitude faster than a full scale \gls{tddft}.
This enables not only rapid screening of configuration space but computations for much larger systems than those accessible by fully-fledged \gls{tddft} calculations on complete systems (or other first-principles approaches), while still maintaining the underlying accuracy of the individual coupled elements.
It thereby becomes possible to study, e.g., ensembles of multiple \glspl{np}, mixtures of multiple \glspl{np} interacting with multiple molecules or aggregates, and -- by extension of the formalism -- also the properties of such systems inside of cavities.

The approach employed here has a long history primarily in the context of molecular aggregates \cite{DeV64, FidKnoWie91}.
Here, we have demonstrated that it is equally applicable to \gls{np} and \gls{np}-molecule assemblies.
More importantly it is shown that by using dynamic polarizabilities from first-principles calculations near-quantitative predictions become possible also for strongly coupled systems.

We note that calculations via the \gls{dc} approximation are complementary to electrodynamics simulations using either the finite-difference time-domain method or the discrete dipole approximation, where the latter bears some technical similarities to the present approach.
Specifically, \gls{dc} calculations enable one to address systems with units in the nanometer size range that are commonly difficult or impossible to access using classical electrodynamics.

As noted the \gls{dc} approximation as used here has shortcomings that need to be addressed in the future.
In particular it will break at short \gls{np}-\gls{np} or \gls{np}-molecule distances, where higher-order multipoles and charge transfer effects become important.
One should note, however, that particles in solution are usually protected by a ligand shell, which prevents very short particle-particle distances.
In the future the approach could therefore be extended to include higher-order multipoles and to account for ligand shells around particles.
Since systems of interest often are composed of particles in solution, the dielectric screening due to the solvation medium could also be included in the approach, which can be accomplished via polarizable continuous medium \cite{BarCosTom97} approaches commonly used in quantum chemistry methods.
As a final note, it is also of interest to include retardation effects, to accurately describe large systems, and to allow for periodic systems.

\section*{Supplementary Material}
See supplementary material for details on the fitting of the spectra to the coupled oscillator model.

\section*{Acknowledgments}
We gratefully acknowledge the Knut and Alice Wallenberg Foundation (2019.0140, J.F., P.E.), the Swedish Research Council (2015-04153, J.F., P.E.), the Academy of Finland (332429, T.P.R; 295602, M.K.), and the Polish National Science Center (2019/34/E/ST3/00359, T.J.A.).
The computations were enabled by resources provided by the Swedish National Infrastructure for Computing (SNIC) at NSC, C3SE and PDC partially funded by the Swedish Research Council through grant agreement no. 2018-05973 as well as by the CSC -- IT Center for Science, Finland, by the Aalto Science-IT project, Aalto University School of Science, and by the Interdisciplinary Center for Mathematical and Computational Modeling, University of Warsaw (Grant~\#G55-6).

\section*{Data Availability}
This study used the \gls{tddft} data published in Ref.~\onlinecite{RosSheErh19Data}.
The new data generated in this study are openly available via Zenodo at \url{http://doi.org/10.5281/zenodo.4095510}, Ref.~\onlinecite{FojRosAnt21Data}.

\section*{Software used}
The \gpaw{} package \cite{MorHanJac05, EnkRosMor10} with \gls{lcao} basis sets \cite{LarVanMor09} and the \gls{lcao}-\gls{rttddft} implementation \cite{KuiSakRos15} was used for the \gls{rttddft} calculations.
The \nwchem{} suite \cite{ValBylGov10} was used for Casida \gls{lrtddft} calculations.
The \textsc{ase} library \cite{LarMorBlo17c} was used for constructing and manipulating atomic structures.
The NumPy \cite{Numpy2020}, SciPy \cite{Scipy2020} and Matplotlib \cite{Hun07} Python packages and the VMD software \cite{HumDalSch96, Sto98} were used for processing and plotting data.
The emcee \cite{ForHogLan13} Python package was used to fit spectra to the coupled oscillator model.

\end{document}